\documentclass[prb,letterpaper,twocolumn,aps,floatfix]{revtex4-2}
\usepackage{graphicx}
\usepackage{xcolor}
\usepackage{amsmath}
\usepackage{amsfonts}
\usepackage[small,bf]{subfigure}
\newcommand{\beq}{\begin{equation}}
\newcommand{\eeq}{\end{equation}}
\newcommand{\bk}{{{\bf{k}}}}

\newcommand{\br}{{{\bf{r}}}}
\newcommand{\bR}{{{\bf{R}}}}

\newcommand{\bG}{{{\bf{G}}}}

\newcommand{\bB}{{\bf{B}}}

\newcommand{\bq}{{\bf{q}}}

\newcommand{\bb}{{\bf{b}}}

\newcommand{\bj}{{\bf j}}

\newcommand{\beqa}{\begin{eqnarray}}
\newcommand{\eeqa}{\end{eqnarray}}
\newcommand{\pdg}{{\vphantom \dag}}
\newcommand{\dg}{{\dag}}
\newcommand{\bnabla}{{\boldsymbol \nabla}} 
\newcommand{\bsigma}{{\boldsymbol \sigma}}

\newcommand{\ra}{\rightarrow}

\newcommand{\cD}{{\cal D}}

\begin{document}
\title{Chiral charge conservation and ballistic magnetotransport \\ in a disordered Weyl semimetal}
\author{A.A. Burkov}
\affiliation{Department of Physics and Astronomy, University of Waterloo, Waterloo, Ontario 
N2L 3G1, Canada} 
\affiliation{Perimeter Institute for Theoretical Physics, Waterloo, Ontario N2L 2Y5, Canada}
\date{\today}
\begin{abstract}
We demonstrate that in an ideal Weyl semimetal, in which the Fermi energy coincides with the band-touching nodes, weak direct inter-nodal scattering is irrelevant and,
as a result, the chiral charge is (almost) exactly conserved. 
This leads to an experimentally-observable effect: in an applied magnetic field, the charge transport along the field becomes purely ballistic, with the conductance given 
by $e^2/h$ per magnetic flux quantum through the sample cross-section. This is the strongest experimental manifestation of nontrivial topology in Weyl and Dirac semimetals. 
\end{abstract}
\maketitle
\section{Introduction}
\label{sec:1}
It has recently been understood that topology plays an important role not just in the physics of insulators, but in metals as well~\cite{Volovik03,Volovik07,Murakami07,Wan11,Burkov11-1,Burkov11-2}. 
In fact, there exists a deep connection between the two, which may be understood using the concept of ``unquantized" anomalies of emergent 
low-energy symmetries~\cite{Song21,Else21,Gioia21,Wang21,Wang24,Hughes24}. 
Low-energy excitation spectrum of any metallic system may be characterized by an emergent conservation law. 
In ordinary metals this is the well-known emergent property of a Fermi liquid: conservation of the particle number at every point on the Fermi surface. 
Similarly, in topological semimetals, there is separate conservation of the particle number at every point on the gapless manifold in momentum space. 
In particular in point-node semimetals, such as Weyl, this corresponds to emergent conservation of the chiral charge. 
Such emergent symmetries are typically ``anomalous", which means that, if they are interpreted literally as physical ``on-site" symmetries, they may only exist 
on the surface of a higher-dimensional symmetry protected topological (SPT) insulator, which provides a connection between the topology of insulators~\cite{Hasan10,Qi11}
and that of metals. 

The way that anomalous surface states of these abstract higher-dimensional SPT insulators actually get realized as metals in physical dimensions, is that 
the emergent symmetries, which act as on-site internal symmetries at low-energies (IR), are in fact descendants of microscopic (UV) symmetries, which are 
non-on-site crystalline symmetries, such as translations, rotations, mirror, etc. 
This immediately raises the question to what extent these topological concepts survive the introduction of disorder, which generally violates all the crystalline symmetries. 
It turns out that even only average symmetry is often enough to preserve nontrivial topology in metals~\cite{Ma22,Yi24,Yi25}, 
which is analogous to what was found earlier in the context of topological insulators~\cite{Fu12,Mong12}. 
In particular, we have demonstrated that Weyl semimetals are protected against Anderson localization due to the nontrivial topology, along with average translational 
symmetry~\cite{Yi24,Yi25}. 

In this paper we take this a step further and show that topology and average translational symmetry lead to an almost exact (the precise meaning of this is described below) chiral charge conservation in ideal Weyl semimetals, in which the Fermi energy exactly coincides with the Weyl nodes. 
This, in turn, leads to experimentally-observable consequences. Namely, in the presence of an external magnetic field, the longitudinal conductance is dominated 
by a purely ballistic, non-dissipative contribution, given by the conductance quantum $e^2/h$ per every magnetic flux quantum, penetrating the sample cross-section. 
This effect may already be observable in currently existing materials, in particular the Cr-doped Bi$_2$Te$_3$ of Ref.~\cite{Belopolski25}, if effort is made to reduce 
disorder (see below for an extended discussion). 

The rest of the paper is organized as follows. 
In Section~\ref{sec:2} we introduce a low-energy model of a magnetic Weyl semimetal with a pair of nodes in the presence of two different kinds of disorder potential: 
scalar disorder, which only scatters electrons within the same node (intra-nodal scattering) and inter-nodal scattering, which acts as a random Dirac mass. 
We discuss the chiral anomaly, characterizing this Weyl semimetal as a 't Hooft anomaly of its emergent low-energy symmetry, namely the separate conservation of 
the left- and right-handed charges. 
In Section~\ref{sec:3} we discuss the effects of disorder in this model, ignoring the topological effects of the chiral anomaly. 
We demonstrate that in this case, the random Dirac mass generally leads to a finite chiral charge relaxation rate. 
In Section~\ref{sec:4} we show that the chiral anomaly invalidates the conclusions of Section~\ref{sec:3} and results in the irrelevance of direct inter-nodal disorder 
scattering and an almost exact conservation of the chiral charge, as long as the crystal translational symmetry of the clean Weyl semimetal is restored after disorder averaging. This leads to an almost vanishing chiral charge relaxation rate even in a disordered Weyl semimetal with inter-nodal scattering, as long as the Fermi energy coincides with the Weyl nodes. 
In Section~\ref{sec:5} we demonstrate that this leads to experimentally-observable phenomena: the sample conductance in the presence of an external magnetic field 
is ballistic and nondissipative and is given by the conductance quantum per every magnetic flux quantum, penetrating the sample cross-section. 
In Section~\ref{sec:6} we point out that the vanishing chiral charge relaxation rate is a unique feature of a 3D Weyl semimetal, which has to do with the chiral anomaly, 
and demonstrate that in an analogous system in 2D, i.e. a Dirac semimetal with a pair of nondegenerate band-touching nodes, such an effect is absent. We then conclude with a brief summary of our results. 
\section{Model of a disordered Weyl semimetal and other preliminaries}
\label{sec:2}
We will adopt a simple model of a Weyl semimetal with a pair of nodes, separated in momentum space, as proposed theoretically in Ref.~\cite{Burkov11-1} and 
realized experimentally in Ref.~\cite{Belopolski25}. 
Our results, however, are applicable to any ideal Weyl semimetal, which has no trivial states at the Fermi energy. 
The Hamiltonian is given by
\beq
\label{eq:1}
H = \sum_\bk v_F c^\dg_\bk \tau^z \bsigma \cdot \bk\,  c^\pdg_\bk. 
\eeq
Here the eigenvalues of the Pauli matrix $\tau^z$ label the chiralities of the two Weyl nodes, which are implicitly understood to be located 
at two different momenta in the first Brillouin zone (BZ) and $\bsigma$ are the spin Pauli matrices. We will use $\hbar = c = 1$ units throughout this 
paper, except in some of the final results. 
To this we add a disorder potential, which we assume to consist of two distinct contributions. Namely, a scalar potential, which is independent of 
either chirality or pseudospin and only scatters electrons within each node, and an inter-node scattering potential, which may be viewed as a random mass 
term for the pair of massless Weyl fermions in Eq.~\eqref{eq:1}. The random mass, which in essence is the Fourier component of the disorder potential at the wavevector, 
connecting the nodes, is complex and has a magnitude and a phase. The total Hamiltonian is then given by
\beqa
\label{eq:2}
H&=&\int d^3 r \psi^\dg(\br) \left[- i v_F \tau^z \bsigma \cdot \bnabla + \frac{1}{2} m(\br) \tau^+ + \frac{1}{2} m^*(\br) \tau^- \right. \nonumber \\
&+&\left. V(\br) \right] \psi^\pdg(\br). 
\eeqa
We will take both disorder potentials to have zero mean $\langle V(\br) \rangle = \langle m(\br) \rangle = 0$. 
In addition, we will assume gaussian distribution with variance $\gamma^2$  for the scalar disorder potential
\beq
\label{eq:3}
\langle V(\br) V(\br') \rangle = \gamma^2 \delta(\br - \br'). 
\eeq

When the random mass $m(\br)$ is set to zero, i.e. in the absence of any inter-nodal scattering, Eq.~\eqref{eq:2} has a continuous symmetry
\beq
\label{eq:5}
\psi(\br) \ra \psi(\br) e^{i \tau^z \theta}, 
\eeq
which corresponds to separate conservation of the left- and right-handed charge. 
Physically, this continuous symmetry is an emergent (or emanant in modern terminology) IR descendant of a true microscopic UV symmetry, which in the case of Weyl semimetals is the crystal translational symmetry. 

Suppose the Weyl nodes are located at momenta $\bk_{R, L} = \pm Q \hat z$ in the first BZ. Then, crystal translational symmetry acts on the right- and left-handed 
states at the Weyl node momenta as
\beq
\label{eq:6}
\psi_R(\br) \ra \psi_R(\br) e^{i Q a n}, \,\, \psi_L(\br) \ra \psi_L(\br) e^{- i Q a n}, 
\eeq
where $a$ is the lattice constant and $n$ is an integer, which corresponds to Eq.~\eqref{eq:5} with $\theta = \theta_n  = Q a n$. 
If $Q = 2 \pi q  / N a$, where $q \in [-N/2, N/2)$ is an integer and $N$ is the number of unit cells in the $z$-direction, 
then $\theta_n = 2 \pi n q/N$. This means that, for a generic $Q$, the translational symmetry, while discrete, may approximately be viewed as 
a continuous $U(1)$ internal symmetry of the chiral low-energy states. 

By analogy with the $U(1)$ electric charge conservation symmetry and the corresponding electromagnetic gauge fields, it is useful to 
view dimensionless Weyl node momenta $\pm Q a$ as translation symmetry charges, while gradients of $\theta_n/ Q a$ as the corresponding gauge fields.
More precisely, this idea may be formalized starting from the dual description of a crystal in terms of intersecting families of crystal planes, rather than Bravais 
lattice points. 
Recalling the definition of reciprocal lattice vectors as solutions of the equation $e^{i \bG \cdot \bR} = 1$, we may define a possibly deformed crystal in 
terms of intersecting families of crystal planes, given by the solutions of the equation~\cite{Volovik19,Nissinen21}
\beq
\label{eq:7}
\theta^i(\br, t) = 2 \pi n^i, 
\eeq
where $i = 1, 2, 3$ labels the three families of crystal planes, needed to define a 3D crystal and $n^i \in \mathbb{Z}$. 
In an undeformed crystal the phase functions are given by
\beq
\label{eq:8}
\theta^i(\br, t) = \bb^i \cdot \br, 
\eeq
where $\bb^i$ are the primitive translation vectors of the reciprocal lattice, such that $b^i_{\mu} = \partial_{\mu} \theta^i$. 
Even in a distorted crystal 
\beq
\label{eq:9}
e^i_{\mu} = \frac{1}{2 \pi} \partial_{\mu} \theta^i, 
\eeq
may be viewed as local basis vectors of the reciprocal space ($1/2 \pi$ factor has been included in the definition for convenience). 
In addition, the index $\mu$ may also include time, introducing a temporal component of the dual vector $e^i$, which also has a physical meaning. 

The one-form $e^i = \frac{1}{2 \pi} d \theta^i$ may be viewed as a translation gauge field, in the sense that in encodes topological properties of the crystal, 
which may be used to define mixed crystal symmetry-electromagnetic topological responses, in cases where purely electromagnetic responses may not be 
topological. 
Indeed the two-form $d e^i$ determines the dislocation density (i.e. translation symmetry flux), while 
\beq
\label{eq:10}
\oint e^i = N^i, 
\eeq
where $N^i$ is the number of unit cells of a crystal with periodic boundary conditions in the $i$th direction. 
This may be used to define a response to inserting an extra crystal plane, which may also be thought of as translation symmetry flux. 
Note that $e^i$ should be viewed as a discrete gauge field, since all its fluxes are integer-valued. 

Now we may use the concept of the translation gauge field to express the following crucial property of any Weyl semimetal. 
In the low-energy model Eq.~\eqref{eq:2} with $m(\br) = 0$ the translational symmetry acts as an axial $U(1)$ symmetry. 
Such a $U(1)$ symmetry is, however, anomalous. This means that gauging this symmetry generates a topological term
\beq
\label{eq:11}
S = \frac{i \lambda e^2}{4 \pi} \int e^z \wedge A \wedge d A, 
\eeq
where $\lambda = 2 Q a/ 2 \pi$ is the ratio of the Weyl node separation to the magnitude of the primitive reciprocal lattice vector (size of the BZ) in the 
same direction and $A$ is the electromagnetic $U(1)$ gauge field. 
This expresses two defining topological properties of a magnetic Weyl semimetal: Hall conductance per atomic plane $\sigma_{xy} = \lambda e^2/ 2 \pi$ and 
a 1D metal on a magnetic flux line with Luttinger parameter $\lambda$~\cite{Gioia21,Wang21}. 
\section{Disorder averaging and chiral charge relaxation}
\label{sec:3}
Let us now return to Eq.~\eqref{eq:2} and consider the effect of disorder. 
In particular, the main issue we will be interested in here is the chiral charge relaxation, which arises from the 
random Dirac mass term. 
It is convenient to represent the complex random mass in terms of its magnitude and phase
\beq
\label{eq:12}
m(\br) = m e^{i \theta(\br)}. 
\eeq
The magnitude $m > 0$ may approximately be taken to be constant (equal to its average value), while the phase $\theta$ is random and may take any 
value between $0$ and $2 \pi$ (we will ignore the discreteness of the crystal translational symmetry here). 
Making a chiral gauge transformation
\beq
\label{eq:13}
\psi_R \ra \psi_R e^{i \theta/2}, \,\, \psi_L \ra \psi_L e^{- i \theta/2},
\eeq
the Hamiltonian is transformed to 
\beqa
\label{eq:14}
H&=&\int d^3 r \psi^\dg(\br) \left[- i v_F \tau^z \bsigma \cdot \bnabla + \frac{v_F}{2} \bnabla \theta \cdot \bsigma + m \tau^x \right. \nonumber \\
&+&\left. V(\br) \right] \psi^\pdg(\br), 
\eeqa
i.e. the gradient of the phase of the random mass enters into the transformed Hamiltonian as a chiral gauge field. 
The problem thus maps onto a massive 3D Dirac fermion with scalar and chiral vector potential disorder. 

It is instructive to first proceed naively, ignoring the nontrivial topology in the form of the chiral anomaly Eq.~\eqref{eq:11}. 
In this case, assuming a slowly-varying inter-nodal scattering potential, it seems reasonable to simply ignore the phase gradient 
term in Eq.~\eqref{eq:14}, which gives
\beq
\label{eq:15}
H = \int d^3 r \psi^\dg(\br) \left[- i v_F \tau^z \bsigma \cdot \bnabla + m \tau^x + V(\br) \right] \psi^\pdg(\br), 
\eeq
i.e. a massive Dirac fermion with scalar disorder. 

This problem may be treated by the standard methods of diagrammatic perturbation theory. 
The energy eigenstates of the massive Dirac Hamiltonian without disorder in Eq.~\eqref{eq:15} are given by
\beq
\label{eq:16}
\epsilon_{s t}(\bk) = t \sqrt{v_F^2 \bk^2 + m^2} \equiv t \epsilon_\bk, 
\eeq
where $s, t = \pm$ are the label the four pairwise-degenerate eigenstates of the massive 3D Dirac Hamiltonian. 
The corresponding eigenvectors may be expressed as
\beq
\label{eq:17}
|z^{s t}_\bk \rangle = |u^s_\bk \rangle \otimes |v^{s t}_\bk \rangle, 
\eeq
where 
\beqa
\label{eq:18}
|u^s_\bk \rangle&=&\left(\sqrt{\frac{1}{2}\left(1 + \frac{s k_z}{k}\right)}, s e^{i \varphi} \sqrt{\frac{1}{2}\left(1 - \frac{s k_z}{k}\right)}\right)^T, \nonumber \\
|v^{s t}_\bk \rangle&=&\left(\sqrt{\frac{1}{2}\left(1 + t s \frac{v_F k}{\epsilon_\bk}\right)}, t \sqrt{\frac{1}{2}\left(1 - t s\frac{v_F k}{\epsilon_\bk}\right)}\right)^T, \nonumber \\
\eeqa
and $e^{i \varphi} = (k_x + i k_y)/\sqrt{k_x^2 + k_y^2}$. 

Within the self-consistent Born approximation (SCBA), the retarded impurity scattering self-energy satisfies the equation
\beq
\label{eq:19}
\Sigma^R_{s t}(\bk, \omega) = \gamma^2 \sum_{s' t'} \int \frac{d^3 k'}{(2 \pi)^3} |\langle z^{s t}_\bk | z^{s' t'}_{\bk'} \rangle |^2 G^R_{s' t'}(\bk', \omega), 
\eeq
where 
\beq
\label{eq:20}
G^R_{s t}(\bk, \omega) = \frac{1}{\omega - t \epsilon_\bk - \Sigma^R_{s t}(\bk, \omega)}, 
\eeq
is the retarded disorder-averaged Green's function. 

After the standard identification of the imaginary part of the self-energy with the impurity scattering rate
\beq
\label{eq:21}
\textrm{Im} \Sigma^R_{s t}(\bk, \omega) = - \frac{1}{2 \tau}, 
\eeq
we obtain the following self-consistent equation for the scattering rate
\beq
\label{eq:22}
\gamma^2 \int \frac{d ^3 k}{(2 \pi)^3} \frac{1}{\epsilon_{\bk}^2 + \frac{1}{4 \tau^2}} = 1. 
\eeq
This has a nontrivial solution when $\gamma > \gamma_c$, which is given by 
\beq
\label{eq:23}
\gamma_c = \sqrt{\frac{2 \pi^2 v_F^2}{\Lambda \left[1 - \frac{\textrm{arctan}(\Lambda v_F/m)}{\Lambda v_F/m} \right]}} \approx \sqrt{\frac{2 \pi^2 v_F^2}{\Lambda}},
\eeq
where $\Lambda \sim 1/a$ is the upper momentum cutoff and the last approximate equality assumes $m / \Lambda v_F \ll 1$. 
When $\gamma < \gamma_c$ the scattering rate and the density of states at the Fermi energy are zero and the system is a gapped insulator. 
For $\gamma > \gamma_c$, there is a finite disorder-induced density of states at the Fermi energy and diffusive transport at long distances and 
long times. 

The long-distance transport properties are described by the propagator of the diffusion modes, which may be viewed as Goldstone modes, arising 
from the spontaneous breaking of the retarded/advanced symmetry by the nonzero SCBA scattering rate~\cite{Altland_Simons}
\beq
\label{eq:24}
\cD(\bq, \Omega) = [1 - I^{RA}(\bq,\Omega)]^{-1}, 
\eeq
where 
\beqa
\label{eq:25}
&&I^{R A}_{a b}(\bq, \Omega) = \frac{\gamma^2}{4} \Gamma^a_{\alpha_2 \alpha_1} \Gamma^b_{\alpha_3 \alpha_4} \nonumber \\
&\times&\int \frac{d^3 k}{(2 \pi)^3} G^R_{\alpha_1 \alpha_3}\left(\bk + \frac{\bq}{2}, \Omega\right) G^A_{\alpha_4 \alpha_2}\left(\bk - \frac{\bq}{2}, 0\right). 
\eeqa
Here $\Gamma = \sigma \otimes \tau$ are $4 \times 4$ gamma matrices and 
\beq
\label{eq:26}
G^R_{\alpha_1 \alpha_2} (\bk, \omega) = \sum_{s t} \frac{z^{s t}_{\bk \alpha_1} z^{s t *}_{\bk \alpha_2}}{\omega - t \epsilon_\bk + \frac{i}{2 \tau}}. 
\eeq

Our system has two soft modes, related to the diffusion of electric and chiral charge. 
The corresponding gamma matrices are $\Gamma^0 = \sigma^0 \otimes \tau^0$ and $\Gamma^5 = \sigma^0 \otimes \tau^z$. 
Focusing on the chiral charge part of the inverse diffusion propagator $\cD^{-1}_{5 5}$, a straightforward calculation gives 
\beq
\label{eq:27}
\cD^{-1}_{5 5}(\bq, \Omega) \approx - i \Omega \tau + D \bq^2 \tau + \tau/\tau_5, 
\eeq
where $D$ is the diffusion constant and the chiral charge relaxation rate $1/\tau_5$ is given by
\beq
\label{eq:28}
\frac{1}{\tau_5} = \frac{2 m^2 \gamma^2}{\tau} \int \frac{d^3 k}{(2 \pi)^3} \frac{1}{(\epsilon_\bk^2 + \frac{1}{4 \tau^2})^2} = 
\frac{m^2 \gamma^2}{4 \pi v_F^2 \tau \sqrt{m^2 + 1/4 \tau^2}}. 
\eeq
This gives a finite chiral charge relaxation time when $m > 0$, as long as $\gamma > \gamma_c$. 
This result is in agreement with the expectation that, in the presence of inter-nodal disorder scattering the chiral charge 
will not be conserved, as it is in a clean Weyl semimetal. 
\section{Random Dirac mass and the chiral anomaly}
\label{sec:4}
Even though this expectation appears sensible, it is actually incorrect. 
First, the applicability of SCBA to the problem of disordered Weyl fermions in 3D is questionable due to nonperturbative rare region effects, 
which likely lead to a finite density of states and diffusive transport at arbitrarily weak scalar disorder~\cite{Nandkishore14,Pixley15,Pixley16,Yi25}.
This, however, is not a strong objection, since Eq.~\eqref{eq:28} may still be applicable, assuming $1/\tau$ is finite at any disorder strength $\gamma$. 
More importantly, the above considerations ignore (also nonperturbative) effects of the chiral anomaly, which implies that the term $\bnabla \theta \cdot \sigma$ 
in Eq.~\eqref{eq:14} can never be ignored, since, as a chiral gauge field, it couples to the electromagnetic field via the topological term Eq.~\eqref{eq:11} as
\beq
\label{eq:29}
S = \frac{i e^2}{8 \pi^2} \int d \theta \wedge A \wedge d A.
\eeq
As a consequence of such anomaly-induced coupling, topological defects in $\theta$, i.e. vortex lines, carry chiral fermion modes. 
Indeed, consider a vortex line in $\theta$, running along the $x$-direction, such that on the line $d d \theta = \pm 2 \pi$. 
In the presence of such a vortex line, Eq.~\eqref{eq:29} is not gauge invariant under $A \ra A + d \varphi$ 
\beq
\label{eq:30}
S \ra S \pm \frac{i e^2}{2 \pi} \int \varphi (\partial_{\tau} A_x - \partial_x A_0). 
\eeq
This anomaly (failure of gauge invariance) is cancelled by a 1D chiral mode, bound to the vortex line~\cite{CallanHarvey}.
Another, more intuitive, way to arrive at this result is to realize that a vortex line in $\theta$ corresponds to a dislocation line of the charge density wave (CDW), induced 
by the inter-nodal coupling with a specific value of $\theta$. 
Since every 2D plane of the CDW superlattice carries a quantum of Hall conductance $e^2/h$, such a dislocation line must carry a single chiral mode, which is simply 
the edge state of such 2D quantum Hall insulator. 
If we assume that the random mass $m(\br)$ averages to zero on long length scales, restoring the broken translational symmetry at short length scales, such vortex lines with gapless chiral modes must percolate through the entire system~\cite{Ludwig94}. 

Now let us first turn off the scalar disorder and leave only the random Dirac mass. In this case, transport can only happen through the percolating cluster of 1D chiral 
modes, bound to the vortex lines of~$\theta$. 
Consider a single straight vortex line of unit positive vorticity, which we will orient along the $z$-axis for convenience. 
Solving the Dirac equation with such a vortex configuration in the mass, one obtains the wavefunction of the bound state~\cite{CallanHarvey}
\beq
\label{eq:31}
\Psi_+(\br) = e^{- \frac{1}{v_F} \int_0^r d r' |m(r')|} (1, 0, 0, -i)^T, 
\eeq
where the integral is along the radial direction, normal to the vortex line. 
The eigenstate energy of this state is $\epsilon_+(\bk) = v_F k_z$, i.e. this is a 1D right-handed chiral mode. 
For a vortex with opposite vorticity one gets
\beq
\label{eq:31}
\Psi_-(\br) = e^{- \frac{1}{v_F} \int_0^r d r' |m(r')|} (0, 1, -i, 0)^T, 
\eeq
which has eigenstate energy $\epsilon_-(\bk) = - v_F k_z$, i.e. this is a left-handed 1D mode. 
Note that the two states are related to each other as
\beq
\label{eq:32}
\Psi_-(\br) = - i \tau^z \sigma^y \Psi_+(\br). 
\eeq
Since $(i \tau^z \sigma^y)^2 = -1$, a closed loop, made of such 1D bound chiral modes, carries a Berry phase of $\pi$ (the dynamical phase is strictly zero since we are 
considering bound states at zero energy). 
This implies that the dynamics of the electrons on the percolating cluster of chiral modes is not diffusive, 
since diffusion would require random phases on any closed loop~\cite{Lee85}. 

The percolating chiral modes will inevitably intersect (approach closely) and tunneling between modes of different chirality will take place at these intersection points. 
We note that, in the absence of a random scalar potential, the tunneling process itself does not result in random phases on closed loops, only its magnitude is random. 
Given the non-diffusive nature of the dynamics and the chiral anomaly of Eq.~\eqref{eq:11}, which is unaffected by the random mass 
(chiral anomaly is the source of the gapless chiral modes),
one comes to the conclusion that, at long distances, the tunneling between intersecting chiral modes must recreate the massless Dirac dispersion of Eq.~\eqref{eq:1}
after averaging. 
Another way to put this is that the only state that, in the absence of a finite density of states at the Fermi energy or strong interactions, is consistent with the topological response of Eq.~\eqref{eq:11}, is a Weyl semimetal (see e.g. Ref.~\cite{Yi25} for an explicit construction of a Weyl semimetal dispersion out of an array of coupled chiral modes).

In essence, what this means is that the random Dirac mass is irrelevant. This conclusion also agrees with the simple scaling analysis. 
Indeed, taking the random mass to be gaussian-distributed
\beq
\label{eq:33}
\langle m^*(\br) m(\br') \rangle = \gamma_m^2 \delta(\br - \br'),
\eeq
and counting the scaling dimensions in the imaginary time action 
\beqa
\label{eq:33.1}
S&=&\int d \tau d^3 r \psi^\dg(\br) \left[\partial_{\tau} - i \tau^z \bsigma \cdot \bnabla + \frac{1}{2} m(\br) \tau^+\right. \nonumber \\ 
&+&\left.\frac{1}{2} m^*(\br) \tau^- + V(\br) \right] \psi^\pdg(\br), 
\eeqa
where we have set $v_F = 1$ for simplicity, while using the standard convention that $\textrm{dim}[r] = -1$, one obtains
$\textrm{dim}[\psi] = 3/2$, $\textrm{dim}[m] = 1$, which implies 
\beq
\label{eq:34}
\textrm{dim}[\gamma_m^2] = -1. 
\eeq
This means that the variance of the random mass potential is irrelevant and free massless Dirac dispersion is restored at long distances. 
On its own, this scaling analysis is suspect, since exactly the same arguments would imply irrelevance of weak scalar disorder as well. 
This is quite likely to be incorrect due to the already mentioned nonperturbative rare region effects~\cite{Nandkishore14,Pixley15,Pixley16,Yi25}.
However, in the context of the random mass disorder, the agreement of the scaling analysis with the chiral anomaly considerations,
strongly suggests that in this case it does in fact correspond to reality. 

This is also consistent with the absence of localization in Weyl semimetals (for sufficiently weak disorder), demonstrated in Ref.~\cite{Yi24}. 
Localization in this system can only arise from inter-nodal scattering, since an isolated Weyl node, being an edge state of a 4D quantum Hall insulator, can 
never be localized. The absence of localization, which may be demonstrated using the nonlinear sigma model and without explicitly using the 
random Dirac mass argument, then implies the irrelevance of the random mass.
In this sense, the chiral charge conservation and the lack of localization arise from the same source and are closely related. 

The above analysis and the resulting conclusions are analogous to the 2D Dirac fermion theory of the quantum Hall plateau transition of Ref.~\cite{Ludwig94}. 
This is not accidental, given a close connection between the 3D magnetic Weyl semimetal and 2D quantum Hall plateau transition~\cite{Yi25}, which arises due to the 
fact that one may view the magnetic Weyl semimetal as an intermediate phase between a 3D quantum Hall and ordinary insulators. 
The argument for the (marginal) irrelevance of the random mass in the 2D case is essentially exact due to the existence of an exact mapping between the 2D Dirac fermion with a random mass and the 2D random-bond Ising model, as demonstrated in~\cite{Ludwig94}. Such a mapping does not exist in the 3D case and our arguments are therefore less rigorous and can not be viewed as a proof, but a well-justified conjecture. 

Adding scalar disorder potential will not change the above conclusions significantly. Its main effect will be to randomize the phases, accumulated on closed loops in the 
percolating cluster. This will make the electron dynamics diffusive, likely at even infinitesimally small scalar disorder due to the already mentioned nonperturbative rare region effects~\cite{Nandkishore14,Pixley15,Pixley16,Yi25}. This will not affect the emergent conservation of the chiral charge, which, as implied by the analysis of this section, is always restored upon disorder averaging, even when it is violated at short length scales. 

What does eventually lead to a nonzero chiral charge relaxation are nonlinear corrections to the Weyl dispersion, which invalidate the random mass model of 
Eq.~\eqref{eq:2} at finite energies. Specific details depend on the realization of the Weyl or Dirac semimetal, but what generally happens may be understood qualitatively
using a simple model, in which the random mass term acquires a momentum dependence, which we take to be quadratic
\beq
\label{eq:34.1}
m(\bk) \approx m_0 + m_2 \bk^2. 
\eeq
The parameter $m_2$ may be estimated from the condition $v_F \Lambda \sim m_2 \Lambda^2$, which gives $m_2 \sim v_F/ \Lambda$. 
This vanishes at the Weyl nodes and does not violate translational symmetry, but does induce mixing between the Weyl fermions of different chirality at any finite 
momentum. This, in combination with a nonzero scalar disorder potential, leads to chiral charge relaxation. 

The corresponding contribution to the chiral charge relaxation rate is given by (see Eq.~\eqref{eq:28})
\beq
\label{eq:34.2}
\frac{1}{\tau_5} = \frac{2 \gamma^2}{\tau} \int \frac{d^3 k}{(2 \pi)^3} \frac{m^2(\bk)}{(\epsilon_\bk^2 + \frac{1}{4 \tau^2})^2}. 
\eeq
Evaluating this assuming $\Lambda \ell \gg 1$ gives
\beq
\label{eq:34.3}
\frac{1}{\tau_5} \sim \frac{1}{\tau} \frac{1}{G_0 (\Lambda \ell)^2}, 
\eeq 
where $1/G_0 = \gamma^2/v_F^3 \tau$ is a dimensionless parameter, quantifying the intra-nodal scattering strength. 
The physical meaning of different factors in Eq.~\eqref{eq:34.3} is easy to understand. 
The $1/G_0$ factor is a measure of the intra-nodal scattering strength, while the $1/(\Lambda \ell)^2$ factor is a measure of the degree of chirality-mixing 
nonlinearity of the Weyl fermion spectrum within the $1/\tau$ energy window. 
Both have to be present to relax the chiral charge. 
In the regime when the disorder is not too strong we can expect $G_0 \gg 1$ and $\Lambda \ell \gg 1$, 
which means that the chiral charge coherence time is very long $\tau_5 \gg \tau$. 
In this case there exists a ``mesoscopic" regime, in which the sample size $L$ may be much less than the chiral charge relaxation, or coherence, length $L_5 = \sqrt{D \tau_5}$. In this regime the chiral charge may be taken to be strictly conserved, which leads to experimentally-observable phenomena, described in Section~\ref{sec:5}. 

Eq.~\eqref{eq:34.3} can be expected to hold as long as $\epsilon_F \ll 1/\tau$. 
In the opposite limit, the chiral charge relaxation rate will have a quadratic dependence on the Fermi energy~\cite{Burkov_lmr_prb}.
Similarly at a finite temperature when $T > 1/\tau$, we will have $1/\tau_5 \sim T^2$.

\section{Observable consequences of the chiral charge conservation}
\label{sec:5}
The most important consequence that follows from the analysis of Section~\ref{sec:4} is that the chiral charge conservation is always restored at long distances, as long as the translational symmetry is restored upon disorder averaging. This is important because it has direct experimentally-observable implications. 
To reveal these consequences, consider transport equations for the electric and chiral charges, which follow from the above analysis. 
Chiral anomaly-driven transport phenomena in Weyl and Dirac semimetals have been discussed extensively before~\cite{Spivak12,Burkov_lmr_prb,Altland16,Burkov_gphe,Burkov_Drude,Shovkovy18}, here we will just highlight specifically what happens in the most interesting case with strictly 
conserved chiral charge. 

Since both electric and chiral charges are strictly conserved at the Fermi energy, both satisfy diffusion equations. 
However, the chiral anomaly term, Eq.~\eqref{eq:11} leads to coupling between the electric and chiral charges.  
This follows from the fact that, as discussed in Section~\ref{sec:2}, Eq.~\eqref{eq:11} implies that every magnetic flux line carries a 1D metal. 
One then obtains the following equations~\cite{Burkov_lmr_prb,Altland16,Burkov_gphe,Burkov_Drude}
\beqa
\label{eq:35}
\frac{\partial n}{\partial t}&=&D \bnabla^2 n + \frac{e^2}{2 \pi^2} \bB \cdot \bnabla \mu_5, \nonumber \\
\frac{\partial n_5}{\partial t}&=&D \bnabla^2 n_5 + \frac{e^2}{2 \pi^2} \bB \cdot \bnabla \mu, 
\eeqa
where $n$ and $n_5$ are electric and chiral charge densities respectively, $\bB$ is the applied 
magnetic field, and $\mu$, $\mu_5$ are the electric and chiral electrochemical potentials. 
The densities are related to the corresponding chemical potentials by $n = g \mu$ and $n_5 = g \mu_5$, 
where $g = 1/2 \pi \gamma^2 \tau$ is the disorder-induced density of states at the Fermi energy~\cite{Yi24}. 
From the first of Eqs.~\eqref{eq:35}, the electric current density is given by
\beq
\label{eq:36}
\bj = \frac{\sigma}{e} \bnabla \mu + \frac{e^3}{2 \pi^2} \mu_5 \bB, 
\eeq
where $\sigma = e^2 g D$ is the ordinary Drude conductivity. 

To compute DC transport properties, we need to solve these equations in the steady state with appropriate boundary conditions. 
Let us assume a sample in the form of a cube with a side $L$ with a steady-state current $I$ that flows in the $z$-direction $j_z = I/L^2$. 
Using Eq.~\eqref{eq:36}, we may express the electrochemical potential $\mu$ in terms of $\mu_5$ and the current $I$ as
\beq
\label{eq:37} 
\frac{d \mu}{d z} = \frac{e I}{\sigma L^2} - \frac{\mu_5}{L_a}, 
\eeq
where 
\beq
\label{eq:38}
L_a = \frac{2 \pi^2 \sigma}{e^3 B}, 
\eeq
is a length scale, above which the first-derivative terms in Eqs.~\eqref{eq:35} start dominating the conventional diffusion terms. 
This new length scale was first introduced by Altland and Bagrets in Ref.~\cite{Altland16}.
Substituting this into the second of Eqs.~\eqref{eq:35}, we obtain
\beq
\label{eq:39}
\frac{d^2 \mu_5}{d z^2} - \frac{\mu_5}{L_a^2} = - \frac{e I}{\sigma L_a L^2}. 
\eeq
This equation needs to be supplemented by boundary conditions, which we take to be 
\beq
\label{eq:40}
\mu_5(z = \pm L/2) = 0. 
\eeq
The physical meaning of this condition is that we take the current leads, attached to the sample at $z = \pm L/2$, to be made of ordinary metal, in which 
the chiral charge quickly relaxes to zero. 

The solution of Eq.~\eqref{eq:39} with boundary conditions \eqref{eq:40} is given by
\beq
\label{eq:41}
\mu_5(z) = \frac{e I L_a}{\sigma L^2} \left[1 - \frac{\cosh(z/L_a)}{\cosh(L/2 L_a)}\right]. 
\eeq
Substituting this result back into Eq.~\eqref{eq:37}, we can obtain the voltage, which develops in response to the current $I$
\beq
\label{eq:42}
V = \frac{1}{e} \int_{-L/2}^{L/2} d z \frac{d \mu}{d z} = \frac{2 I L_a}{\sigma L^2} \tanh(L/2 L_a). 
\eeq
This exhibits a nontrivial sample size dependence, which reflects the competition between the standard dissipative diffusive transport, described by the 
second-derivative terms in Eq.~\eqref{eq:35}, and ballistic nondissipative transport due to the chiral anomaly, described by the first-derivative terms. 
At short length scales $L < L_a$, diffusive transport dominates and we obtain the standard Ohmic conductance
\beq
\label{eq:43} 
G = I/V = \sigma L. 
\eeq
At long length scales $L > L_a$, however, the character of the transport changes qualitatively. 
In this case we obtain
\beq
\label{eq:44}
G = \frac{\sigma L^2}{2 L_a} = \frac{e^2}{2 \pi} \frac{e B L^2}{2 \pi} = \frac{e^2}{h} \frac{B L^2}{h c/e}, 
\eeq
where we have restored explicit $\hbar$ and $c$ in the final result. 
Thus the conductance at large length scales becomes nondissipative and ballistic, given by $e^2/h$ per every magnetic flux quantum, 
penetrating the sample cross-section. This is a striking result, which arises from the combination of emergent chiral charge conservation and the chiral anomaly.
This is a new macroscopic quantum transport phenomenon, which is a characteristic feature of ideal Weyl (and Dirac) semimetals. 
When $1/\tau_5$ is finite, this ballistic conductance will be observed in the ``mesoscopic" regime, in which the sample size $L$ satisfies 
$L_a < L < L_5^2/L_a$~\cite{Altland16,Burkov_gphe,Burkov_Drude}.
\section{Discussion and conclusions}
\label{sec:6}
To highlight the nontrivial nature of the emergent chiral charge conservation in Weyl semimetals, it is useful to compare with a very similar system in 2D, 
which is sometimes called ``2D Weyl semimetal"~\cite{2DWeyl}. 
In this case there exists a pair of nondegenerate 2D Dirac points, separated by $2 Q$ in momentum space. This may be viewed as either a 2D analog 
of a Weyl semimetal or spinless graphene. 
The Dirac points are protected by translational symmetry, as well as mirror symmetry in the plane containing the band-touching nodes~\cite{Burkov18-2}. 
The topological response in this case takes the form~\cite{Wang21,Hughes24}
\beq
\label{eq:45}
S = \frac{i \lambda e}{2} \int e^x \wedge d A, 
\eeq
where $\lambda = 2 Q a/ 2\pi$ is the ratio of the Dirac node separation to the primitive reciprocal lattice vector. 
Physically, this describes electric polarization in the direction, perpendicular to the mirror plane. 
The analog of the chiral charge is the ``valley" charge in this case. 

We may now carry out the same analysis of the internodal scattering as in Section~\ref{sec:4}. 
As a consequence of the topological term Eq.~\eqref{eq:45}, the phase $\theta$ of the translation-symmetry breaking mass term will couple to the 
electromagnetic field as
\beq
\label{eq:46}
S = \frac{i}{4 \pi} \int d\theta \wedge d A. 
\eeq
This leads to a half-quantized electric charge in the core of a vortex in $\theta$. 
Restoring the broken translational symmetry at long length scales requires creating such vortices with half-quantized charges in the core. 
However, since the vortices (which are points rather than lines in this case) may be arbitrarily far apart in a macroscopic sample, there is no analog of the percolation of gapless modes, which 
happens in 3D. This means that a random Dirac mass in 2D is relevant (this is a different 2D Dirac mass than the one arising in the theory of the quantum Hall 
plateau transition, which is marginally irrelevant, as discussed in Section~\ref{sec:4}) and leads to both localization and the valley charge relaxation due to 
direct intervalley scattering, in sharp contrast to the 3D case. 

In conclusion, in this paper we have demonstrated that in Weyl semimetals direct inter-nodal scattering is irrelevant, at least when sufficiently weak,
and does not lead to either localization or even relaxation of the chiral charge, as long as the translational symmetry is preserved on average. 
The only contributions to the chiral charge relaxation rate come from nonlinear corrections to the Weyl dispersion along with intra-nodal scattering. 
This leads to extremely long chiral relaxation times when the Fermi energy is aligned with the Weyl nodes and when the intra-nodal scattering is not 
too strong. This results in striking observable consequences in the ``mesoscopic" regime, in which 
the temperature is low and the sample size is less than the chiral charge relaxation length. In this case the sample magnetoconductance is ballistic, 
given by $e^2/h$ per magnetic flux quantum through the sample cross-section. This regime may be achievable in the currently existing 
materials, in particular in the MBE-grown Cr-doped Bi$_2$Te$_3$ samples of Ref.~\cite{Belopolski25}, if effort is made to reduce disorder and align the 
Fermi energy with the Weyl nodes. If we assume that the Fermi energy is at charge neutrality for the samples studied in Ref.~\cite{Belopolski25}, we 
may estimate the 2D sheet conductivity for a sample of thickness $d$ to be (restoring explicit Planck's constant)~\cite{Altland16}
\beq
\label{eq:47}
\sigma^{2D} \sim \frac{e^2}{h} \frac{d}{\ell}. 
\eeq
Taking $\sigma^{2D} \approx 50 e^2/h$, as reported in Ref.~\cite{Belopolski25}, we obtain $\ell \sim 1 \textrm{nm}$, which is quite short. 
If we take $v_F \sim 10^7 \textrm{cm/s}$, we then obtain $\hbar/\tau \sim 0.1 \textrm{eV}$, which, correspondingly, is very large. 
Even though this estimate should be taken with a grain of salt, since the location of the Fermi energy is not actually known with any certainty, 
it does suggest that the samples are strongly disordered and likely outside of the regime in which the low-field ballistic transport, described in Section~\ref{sec:5},
could be observable.
\begin{acknowledgments}
We acknowledge useful discussions with Roni Ilan and Chong Wang.
Financial support was provided by the Natural Sciences and Engineering Research Council (NSERC) of Canada (computing and manuscript preparation) and the Center for Advancement of Topological Semimetals, an Energy Frontier Research Center funded by the U.S. Department of Energy Office of Science, Office of Basic Energy Sciences, through the Ames Laboratory under contract DE-AC02-07CH11358. 
Research at Perimeter Institute is supported in part by the Government of Canada through the Department of Innovation, Science and Economic Development and by the Province of Ontario through the Ministry of Economic Development, Job Creation and Trade.
\end{acknowledgments}
\bibliography{references}
\end{document}